\documentclass[preprint,12pt]{aastex}
\usepackage[usenames]{color}
\newcommand{\Hii}{\ion{H}{2}}

\newcommand{\kms}{km s$^{-1}$}
\begin{document}

\title{Discovery of New Interacting Supernova Remnants in the Inner Galaxy}

\author{John W. Hewitt and Farhad Yusef-Zadeh}
\affil{Department of Physics and Astronomy, Northwestern University, Evanston, 
IL 60208}

\begin{abstract}
OH(1720 MHz) masers are excellent signposts of interaction between supernova
remnants(SNRs) and molecular clouds. Using the GBT and VLA we have surveyed 75
SNRs and 6 candidates for masers. Four SNRs are detected
with OH masers: G5.4-1.2, G5.7-0.0, G8.7-0.1 and G9.7-0.0. Two SNRs,
 G5.7-0.0 and G8.7-0.1, have TeV $\gamma$-ray counterparts which may
indicate a local cosmic ray enhancement. It has been noted that maser-emitting
SNRs are preferentially distributed in the Molecular Ring and Nuclear Disk. We use
the present and existing surveys to demonstrate that masers are strongly confined
to within $|l|$ $\le $ 50\degr\ at a rate of 15\% of the total SNR population. All new
detections are within 10\degr\ Galactic longitude emphasizing this trend.
Additionally, a substantial number of SNR masers have peak fluxes at or below the detection threshold of existing surveys. This calls into question whether maser surveys of Galactic SNRs can be considered complete and how many maser-emitting remnants remain to be detected in the Galaxy.
\end{abstract}


\section{Introduction}

Uncovering supernova remnants (SNRs) which are directly interacting with molecular clouds is an important avenue for understanding energy injection and chemical evolution in the Galactic interstellar medium. Yet only a small fraction of known remnants have been identified as interacting. 

Maser emission from the 1720 MHz transition of hydroxyl (OH) accompanied by absorption in the other ground-state lines is a well-established indicator of interaction. SNR-type masers are important probes of the temperature, density, magnetic field, energetic photon flux and shock chemistry arising in the cooling post-shock gas behind the shock front \citep{frail94,wardle02}. They are one of the few reliable methods to determine SNR distances which in turn constrains the remnant's age, size and supernova energy. Early single dish surveys with interferometric follow-up were able to search most of the known population of SNRs identifying 20 Galactic remnants with OH(1720 MHz) masers \citep{frail96,green97}. Since then multi-frequency surveys have nearly doubled the number of known Galactic SNRs \citep{gray94iv,whiteoak96,brogan06}. As has been noted, the inner Galaxy with its large reservoir of dense molecular material is a particularly promising region to search for SNR masers with a high incidence of maser-emitting(ME) SNRs near the Galactic center \citep{green97,fyz99}.

Masers identify a valuable class of high-energy sources. A close association is seen with mixed-morphology SNRs which are thought to arise from thermal conduction of gas into the remnant interior (Yusef-Zadeh et al. 2003). Furthermore, 9 of 24 ME SNRs have coincident EGRET or HESS sources \citep{esposito96,aharonian06}. This may indicate that ME SNRs are sites of cosmic ray acceleration where the presence of dense molecular material acts as a target for cosmic rays producing prominent GeV and TeV $\gamma$-rays \citep{fatuzzo06}. There may also be a causal relationship between mixed-morphology or $\gamma$-ray remnants and OH masers. Increased ionization from either the soft X-ray flux from the interior or a local cosmic ray enhancement can produce the requisite OH abundance in the post-shock gas in which masers form \citep{wardle99}.

Here we report on a new survey for OH(1720 MHz) masers which includes the thirty recently discovered SNRs in the inner Galaxy \citep{brogan06} as well as two HESS sources and 49 SNRs from Green's (2006) Catalog which had not been previously surveyed. We find OH(1720 MHz) masers in four SNRs: G5.4-1.2, G8.7-0.1, G9.7-0.0 and G5.7-0.0 which was previously listed as an unconfirmed candidate SNR. Of the new detections G5.7-0.0 and G8.7-0.1 are coincident with HESS TeV detections and possible sites of cosmic ray acceleration. Given these new detections we re-analyze the properties of ME SNRs as a class in $\S$4. We find that the statistical distribution of maser peak fluxes reported in the literature casts doubt as to whether the sensitivity of previous surveys is sufficient for them to be considered complete. It is likely that deeper searches for OH(1720 MHz) masers will identify further ME SNRs.

\section{Observations}\label{sec:observations}
Motivated by the incompleteness of previous surveys for OH(1720 MHz) emission and the prospect that ME SNRs lie preferentially in the inner Galaxy, we used the Green Bank Telescope(GBT) and Very Large Array(VLA)\footnote{The National Radio Astronomy Observatory is a facility of the National Science Foundation operated under cooperative agreement by Associated Universities, Inc.} to search for OH(1720 MHz) masers as a tracer of interaction. This new survey is comprised of both a VLA survey of 18 SNRs as 
well as a separate GBT survey of 69 targets with VLA follow-up. Table 1 lists all 75 SNRs and 6 candidates with notes as to the nature of the source.

We used the GBT to survey 63 SNRs, four unconfirmed candidate SNRs from Brogan et al. (2006) and two unidentified HESS sources from Aharonian et al. (2006). Multiple GBT pointings were used to map the entirety of SNRs larger than a GBT beam (7\arcmin\ at 1.7 GHz). All four ground state transitions of OH at 1612.231, 1665.4018, 1667.359 and 1720.530 MHz were simultaneously observed, each with a 12.5 MHz bandwidth and a velocity resolution of 0.53 \kms\ channel$^{-1}$. Spectra have an RMS sensitivity of 10--50 mJy with a median of 25 mJy per pointing. Follow-up VLA observations were obtained in the D-array for six GBT detections where the criteria for SNR-type masers was met: narrow 1720 MHz emission line; corresponding OH main-line and 1612 MHz absorption; and within the boundary of the remnant inferred from radio emission. VLA observations obtained a sensitivity of 9 mJy beam$^{-1}$ and confirmed masers in SNRs G5.7-0.0 and G9.7--0.0. A 780 kHz bandwidth with 255 channels was used with 0.5 \kms\ resolution centered at the maser velocity. 

Additionally we used the VLA in C-array to observe eighteen SNRs at 1720 MHz which were not included in the GBT survey. These SNRs were located within -10\degr\ $<$ l $<$ 18\degr . A sensitivity of 7--9 mJy beam$^{-1}$ channel$^{-1}$ was obtained with a 1.56 MHz bandwidth covering velocities from -135 to +135 \kms\ with a resolution of 2.1 \kms\ channel$^{-1}$. SNRs G5.4--1.2 and G8.7--0.1 were later observed with the GBT to confirm OH absorption associated with the VLA detections of 1720 MHz masers. Further survey details will be presented in Hewitt (2009, PhD thesis). 

\begin{deluxetable}{rrl|rrl}
\tablecaption{Observed SNRs and candidates\label{tbl:list}}
\tabletypesize{\scriptsize}
\tablewidth{0pt}
\tablehead{ 
 \colhead{l} & \colhead{b} & \colhead{Notes} &
 \colhead{l} & \colhead{b} & \colhead{Notes} }
\startdata
 4.2 & -3.5 & v &        18.6 & -0.2 &D     \\
 4.8 & +6.2 & v &        18.9 & -1.1 &D     \\
 5.2 & -2.6 & v &        19.1 & +0.2 &D     \\
 5.4 & -1.2 & v &        19.1 & +0.9 &B    \\
 5.5 & +0.3 & D&         20.0 & -0.2 &D     \\
 5.7 & -0.0 & B,D,f &    20.4 & +0.1 &D     \\
 5.9 & +3.1 & v &        21.0 & -0.4 &D     \\
 6.1 & +1.2 & v &        21.5 & -0.1 &D     \\
 6.1 & +0.5 & D&         25.5 & +0.0 &D     \\
 6.3 & +0.5 & B,D &      28.6 & -0.1 &D     \\
 6.5 & -0.4 & D&         29.6 & +0.1 &D     \\
 6.4 & +4.0 & v &        30.7 & +1.0 &D    \\
 7.2 & +0.2 & D&         31.5 & -0.6 &D     \\
 7.7 & -3.7 & v &        36.6 & -0.7 &D     \\
 8.3 & -0.0 & D&         36.6 & +2.6 &     \\
 8.7 & -0.1 & v &        40.5 & -0.5 &D     \\
 8.7 & -5.0 & v &        42.8 & +0.6 &     \\
 8.9 & +0.4 & D&         45.7 & -0.4 &D     \\
 9.7 & -0.0 & D,f &      46.8 & -0.3 &D,f   \\
 9.9 & -0.8 & D&         53.6 & -2.2 &     \\
10.5 & -0.0 & D&         55.0 & +0.3 &     \\
11.1 & -1.0 & D&         55.7 & +3.4 &     \\
11.8 & -0.2 & D&         57.2 & +0.8 &     \\
12.2 & +0.3 & D&         59.8 & +1.2 &     \\
12.5 & +0.2 & D&         63.7 & +1.1 &     \\
12.7 & -0.0 & D&         65.7 & +1.2 &     \\
12.8 & -0.2 & D,H &      67.7 & +1.8 &     \\
12.8 & -0.0 & D&         68.6 & -1.2 &D     \\
14.1 & -0.1 & D&         69.7 & +1.0 &     \\
14.3 & +0.1 & D&         73.9 & +0.9 &     \\
15.1 & -1.6 & v &        76.9 & +1.0 &     \\
15.4 & +0.1 & D,f &      84.9 & +0.5 &D,f   \\
15.5 & -0.1 & D,B &      85.4 & +0.7 &     \\
16.0 & -0.5 & D&         85.9 & -0.6 &     \\
16.2 & -2.7 & v &       130.7 & +3.1 &     \\
16.4 & -0.5 & D&        347.3 & -0.5 &v  \\
16.8 & -1.1 & v &       350.0 & -2.0 &v  \\
17.0 & -0.0 & D&        353.9 & -2.0 &     \\
17.4 & -0.1 & D&        356.2 & +4.5 &     \\
17.8 & -0.7 & D,H,f &   358.0 & +3.8 &v  \\
18.1 & -0.1 & D&         \\
\enddata
\tablecomments{B = SNR candidate from \citet{brogan06}, D = single dish 1720 MHz detection, H = unidentified HESS source, v = only observed with the VLA, f = follow-up observations with the VLA were obtained for these GBT detected SNRs}
\smallskip
\end{deluxetable}

\section{Newly Detected SNR Masers}
We have identified four new ME SNRs in a sample of 76 remnants. In general these remnants have not been well studied and are for the first time identified as interacting. Gaussian fits to the maser position and velocity are given in Table \ref{tbl:detections}. A kinematic distance to the SNR is estimated from the maser velocity\footnote{Here we make the common assumption that the maser is near the SNR's systemic velocity  and use the rotation curve of \citet{fich89} with R$_\odot$ = 8.5 kpc and $\Theta_\odot$ = 220 \kms .}.
All SNRs show symmetric absorption from the 1667, 1665 and 1612 MHz transitions at the position of the maser. This suggests the shock is directed largely across the plane of the sky at the position of the maser \citep{claussen97}.
This shock geometry maximizes the coherent velocity pathlength in the OH gas necessary for masing, and requires the masers to lie near the systemic velocity of the SNR yielding a reliable kinematic distance. We use OH absorption against the bright SNR continuum to resolve the near/far kinematic distance ambiguity.
 
Each SNR is discussed in the order of interest, summarizing the evidence for interaction and placing the new maser detections in context.

\begin{deluxetable}{lllrrrrrrrrrrrrr}
\tablecaption{Detected OH(1720 MHz) Masers toward supernova remnants \label
{tbl:detections}}
\tabletypesize{\scriptsize}
\tablewidth{0pt}
\tablehead{ \colhead{SNR} & \colhead{RA} &  \colhead{DEC} &  \colhead{V$_{LSR}$}&  \colhead{$\Delta$V} & \colhead{Peak Flux} & \colhead{Int. Flux} &   \colhead{RMS} & \colhead{T$_{B}$} & \colhead{d} & \\
\colhead{Name} & \colhead{(J2000)} &  \colhead{(J2000)} &  \colhead{(\kms )} &  \colhead{(\kms )} &   
\colhead{(mJy beam$^{-1}$)} & \colhead{(mJy)} &  \colhead{(mJy beam$^{-1}$)} & 
\colhead{(K)} & \colhead{(kpc)} }
\startdata
5.4-1.2   & 18:01:18.5 & -24:54:58 &--21 &$\le$2&$\ge$61 & 113& 7& $\ge$73  & 5.2 \\
          & 18:01:18.9 & -24:54:20 &--21 &$\le$2&$\ge$42 & 114& 7& $\ge$49  & \\
5.7-0.0   & 17:59:01.6 & -24:04:50 & +13 &0.5   &     59 & 137& 9& $\ge$19  & 3.1,13.7\\
8.7-0.1   & 18:06:18.7 & -21:24:04 & +36 &$\le$2&$\ge$632& 693& 9& $\ge$1870& 4.5 \\
9.7-0.0   & 18:07:53.8 & -20:32:45 & +43 &0.7   &     68 & 92 & 9& $\ge$15  & 4.7
\enddata
\tablecomments{Some masers were not resolved spectrally so only lower limits are placed on the peak flux and upper limits on the line width.}
\smallskip
\end{deluxetable}

\begin{figure}
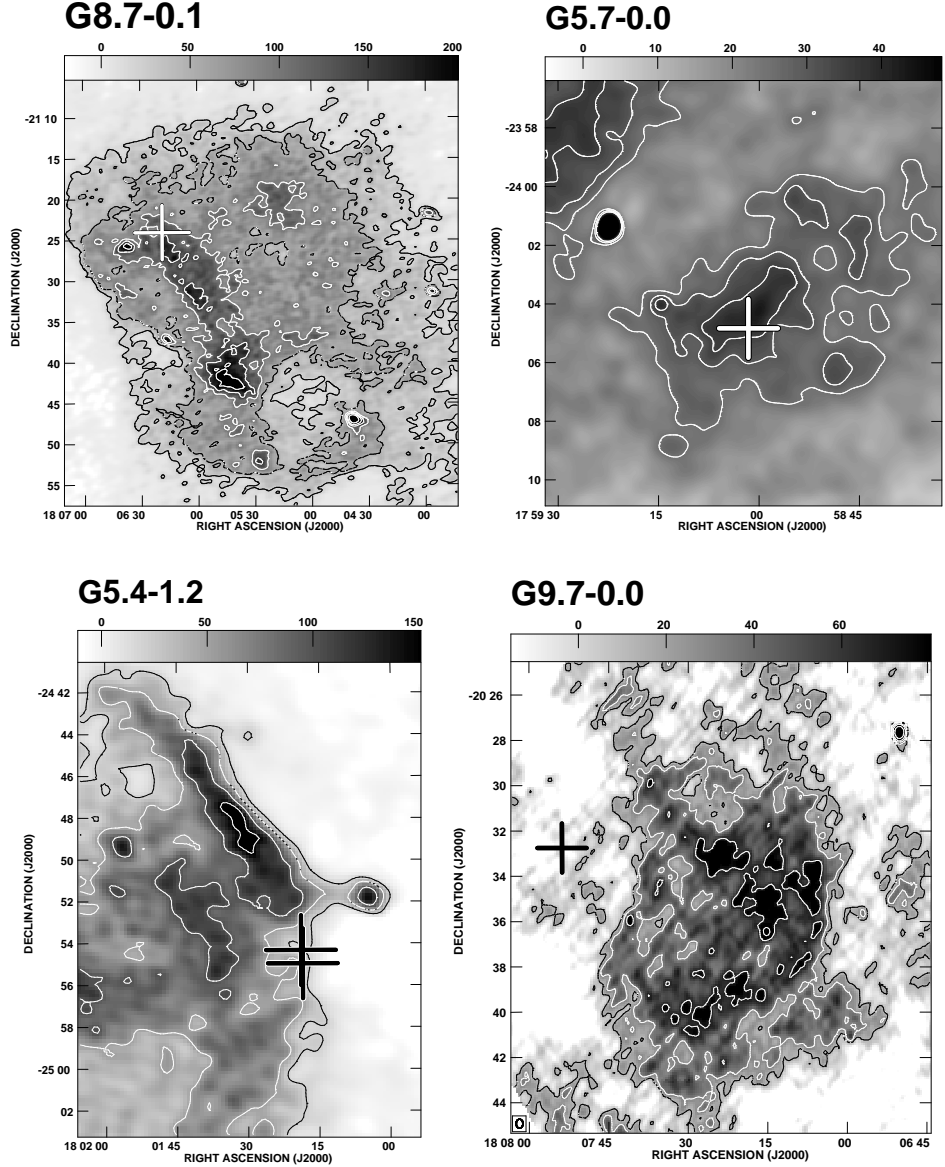

\centering
\includegraphics[height=3in]{f1a.ps}
\includegraphics[height=3in]{f1b.ps}
\includegraphics[height=3.3in]{f1c.ps}
\includegraphics[height=3.3in]{f1d.ps}
\caption{90cm continuum images of newly detected ME SNRs, with the masers demarcated as crosses. From left to right
{\bf G8.7-0.1:} image from  { \citet{kassim90} with a beam of  55\arcsec$\times$37\arcsec. Contours levels at 20, 50, 100, 150 and 200 mJy beam$^{-1}$.\label{fig:8.7-0.1}}
{\bf G5.7-0.0:} image from the Multi-Array Galactic Plane Imaging Survey (MAGPIS) \citet{helfand06} convolved with a 30\arcsec\ beam. Contour levels at 25, 30 and 35 mJy beam$^{-1}$.\label{fig:5.7-0.0}
{\bf G5.4-1.2:} image from \citet{frail94pulsar} with a beam of 55\arcsec$\times$41\arcsec. Contour levels at 20, 50, 100, 150, 200 and 300 mJy beam$^{-1}$.\label{fig:5.4-1.2}
{\bf G9.7-0.0:} MAGPIS image with a beam of 6\arcsec . Contour levels at 12, 30, 70, 150 and 400 mJy beam$^{-1}$. \label{fig:9.7-0.0}
}
\end{figure}

{\smallskip \bf \noindent G8.7--0.1 (W30)} 
SNR G8.7-0.1 is a large 45\arcmin\ remnant which together with nine \Hii\ regions along its south-eastern extent forms the W30 complex \citep{kassim90}. The SNR interior is filled by a thermal X-ray plasma (T$\sim$6$\times$10$^6$ K) marking G8.7-0.1 as a mixed-morphology remnant \citep{finley94}. Extended $\gamma$-ray source HESS J1804-216 is detected along the western shell and may be further evidence of shock interaction and cosmic ray acceleration \citep{aharonian06}\footnote{PSR J1803-2137 also lies in the vicinity of the TeV source making a PWN scenario possible, but the X-ray flux is surprisingly low if $\gamma$-rays are produced via the inverse-compton process \citep{bamba07}.}.

A single, bright OH(1720 MHz) maser is detected at +36 \kms\ along the eastern edge of G8.7-0.1. Figure \ref{fig:8.7-0.1} shows the location of the maser between synchrotron filaments visible at both 20 and 90 cm. Radio recombination line emission from the young HII regions in the vicinity of G8.7-0.1 indicate velocities between +30 and +45 \kms\ \citep{kassim90} similar to that of the SNR. However, we find no compact radio source within 5\arcmin\ of the maser, strengthening our classification as a SNR-type maser. GBT spectra at the position of the maser show symmetric absorption from the main-line and 1612 MHz OH transitions about 10 \kms\ in width. 
Additional absorption features are seen at -29, +4, +15 and +19 \kms\ but not at velocities higher than that of the SNR maser. X-ray observations find a neutral hydrogen column density of only N$_H$$\sim$1.2$\times$10$^{22}$ cm$^{-2}$, so we conclude the SNR G8.7-0.1 lies at a distance of 4.5 kpc.

{\smallskip \bf \noindent G5.7--0.0} 
Identified by \citet{brogan06} as a SNR candidate, SNR G5.7--0.0 is a partial 12\arcmin\ shell with a spectral index of --0.5. Though it is not currently listed in Green's catalog of Galactic SNRs, it is coincident with TeV $\gamma$-ray source HESS J1800-240C \citep{aharonian08w28} and has the characteristics of an interacting SNR which will be an excellent target for further study.

A single OH(1720 MHz) maser is detected at +12.8 \kms\ at a peak in the radio shell as can be seen in Figure \ref{fig:5.7-0.0}. The radio morphology of the SNR is somewhat unclear though given its low surface brightness.
GBT spectra show characteristic OH absorption that is narrow and symmetric. The maser velocity places the remnant at a kinematic distance of either 3.1 or 13.7 kpc. Absorption is seen at +7 and -25 \kms . However a compact radio source also lies within the GBT beam so the near/far distance ambiguity to G5.7-0.0 cannot be resolved with existing observations.

{\smallskip \bf \noindent G5.4--1.2} 
It has long been suspected that SNR G5.4-1.2 is associated with nearby pulsar wind nebula PSR J1801-2451. The pulsar has a characteristic age of 15.5 kyrs at a distance of 5.2$\pm$0.5 kpc. A proper motion limit of 11$\pm$9 mas yr$^{-1}$ was determined for the pulsar, ruling out an origin from the center of SNR G5.4-1.2 within the characteristic age of the pulsar \citep{zeiger08}. However, the radio shell is much brighter along the western edge, suggestive of a density gradient. The current apparent center of G5.4-1.2 may not necessarily be the location of the supernova event. X-ray observations are not sensitive enough to detect thermal emission typical of mixed-morphology remnants \citep{kaspi01}. %Faint H$\alpha$ nebulosity is seen towards the interior of the remnant \citep{zealey79} which has been observed for other interacting remnants.

Two OH(1720 MHz) masers are found along the bright western edge of SNR G5.4-1.2 as seen in Figure \ref{fig:5.4-1.2}. Both lie at a velocity of --21 \kms\ and are located within an arcminute of each other. Absorption at the velocity of the maser appears symmetric with both a narrow and a broad component, $\Delta$V = 3 and 17 \kms\ respectively.
The systemic velocity of the SNR as traced by masers is contrary to rigid Galactic rotation and places G5.4-1.2 within the 3-kpc expanding arm of our Galaxy at a distance of 5.2 kpc \citep{dame08}. HI absorption is seen at +40 to -40 \kms\ towards the SNR indicating a distance of $>$4.5 kpc consistent with our distance determination \citep{frail94pulsar}. Given the similar distance for PSR J1801-2451, only $\sim$3\arcmin\ from the shell of SNR G5.4-1.2, if the pulsar is slightly older than its characteristic age a past interaction with the SNR shock wave cannot be ruled out.

{\smallskip \bf \noindent G9.7--0.0} 
 \citet{frail94pulsar} first identified G9.7-0.0 as a low surface brightness source (10$^{-21}$ W m$^{-2}$ Hz$^{-1}$ sr$^{-1}$ at 327 MHz) with a partial shell 15\arcmin\ in diameter. \citet{brogan06} confirmed the non-thermal nature of the source. 

We detect a single maser at +43 \kms within the projected extent of G9.7-0.0 shown in Figure \ref{fig:9.7-0.0}. GBT spectra toward the position of the maser show absorption at +5, +20 and +43 \kms\ in the 1667, 1665 and 1612 MHz lines. Surveys of CO and HI show emission at velocities up to +160 \kms\ yet no OH absorption features are seen against the SNR at velocities higher than the detected maser. This suggests the SNR lies at the near kinematic distance of 4.7 kpc.

We compared our observations of G9.7--0.0 with Spitzer IRAC and MIPS images at 3.6, 
4.5, 5.8, 8 and 24 \micron\ to see if there was any evidence of an interaction as has been done by 
\citet{reach06}. No IR emission is seen associated with the supernova remnant, but it 
is difficult to detect SNRs against the bright IR background of the Galaxy. 
We note that the 20cm emission is encircled by a series of IR dark clouds. That 
the position of the detected OH(1720 MHz) maser is coincident with one of these clouds is 
suggestive, but further study is needed to confirm that the SNR G9.7-0.0 is an interacting remnant.

\begin{figure}
\centering
a.\includegraphics[width=3in]{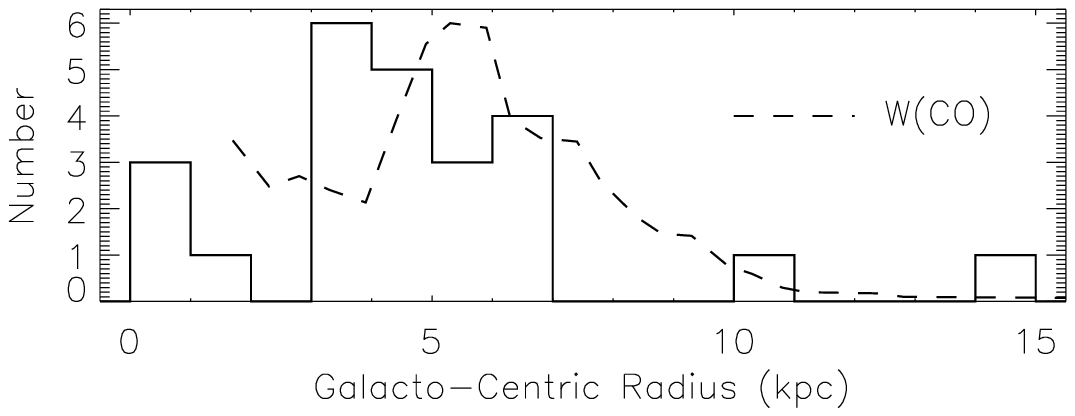}
b.\includegraphics[width=3in]{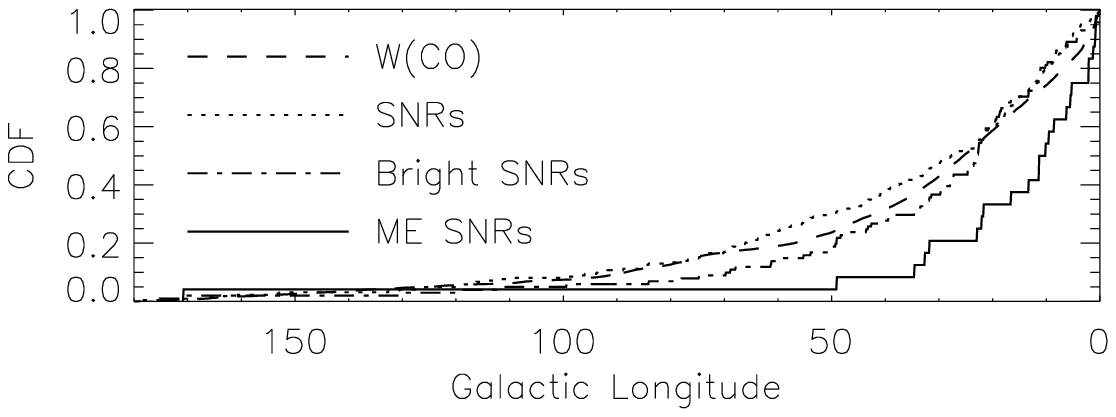}
c.\includegraphics[width=3in]{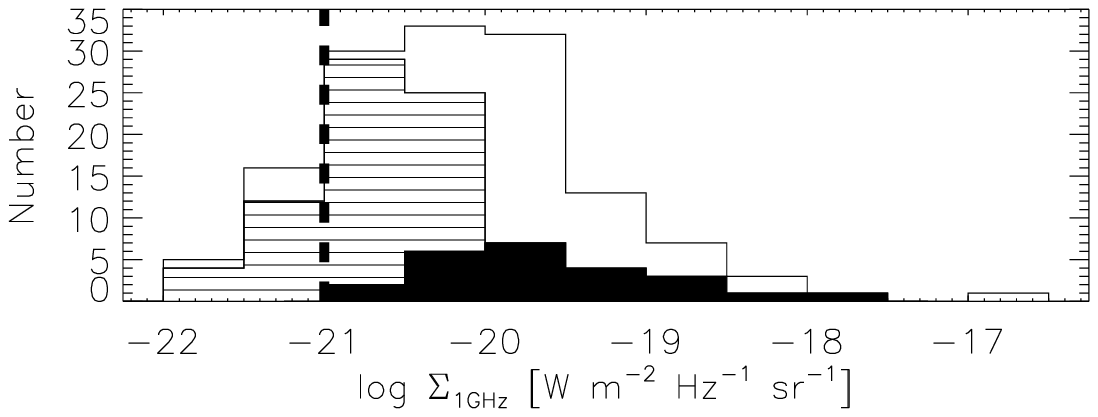}
d.\includegraphics[width=3in]{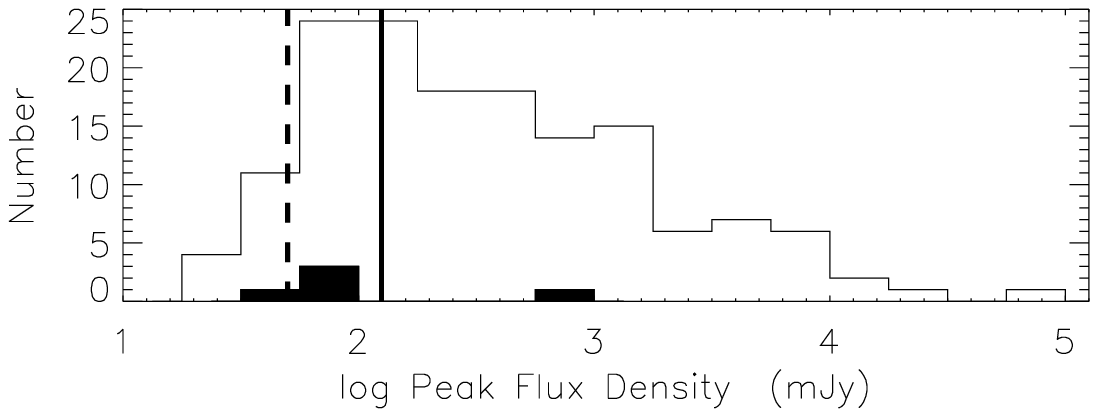}
\caption{Histograms are given which demonstrate the properties of ME SNRs in the Galaxy.
(a) The radial distribution of ME SNRs is compared to the distribution of molecular gas(dashed line) as traced by azimuthally integrated CO emissivity \citet{clemens88} normalized to the peak in ME SNRs.
(b) The cumulative distribution function(CDF) for ME SNRs(solid), all SNRs(dotted), CO emissivity(dashed) and bright SNRs ($\Sigma_{1 GHz} \ge$ 5$\times$10$^{-20}$ W m$^{-2}$ Hz$^{-1}$ sr$^{-1}$, as a function of Galactic longitude. 
(c) The distribution of remnant surface brightness at 1 GHz for three subsets: [empty histogram] SNRs surveyed by Frail et al. (1996) and Green et al. (1997); [hatched histogram] SNRs surveyed by this work; [filled histogram] ME SNRs. A dashed line shows the completeness limit for surface brightness ($\Sigma_{1 GHz}$) in the \citet{brogan06} survey.
(d) The distribution of the peak flux density of all detected SNR masers is given by the unfilled histogram. The new masers identified in this work are given by the filled histogram. Lines are drawn for single dish surveys(solid) and interferometer follow-up(dashed) at the 5$\sigma$ detection limits of 125 and 50 mJy respectively.
\label{fig:histograms}}
\end{figure}

\section{Discussion}
New maser detections presented here make it worthwhile to re-analyze the statistics of ME SNRs. Including this work, 223 of 266 Galactic SNRs have been searched for masers using both single dishes and interferometers with sensitivities of 5--25 mJy and 35--160 mJy respectively \citep{frail94,frail96,green97,koralesky98,fyz95,fyz96,fyz99,sjouwerman08}. This has resulted in the detection of 24 ME SNRs in the Galaxy, consistent with earlier estimates that 10$\%$ of SNRs harbor masers. 

Maser-emitting SNRs trace regions of elevated molecular gas density, particularly the Molecular Ring ($|l|\le$50\degr ) and the Nuclear Disk ($|l|\le$5\degr ) \citep{green97}. The galacto-centric distances for all ME SNRs (derived from maser velocities as in \S 3) can be used to obtain a radial distribution shown in Figure \ref{fig:histograms}a. Two-thirds of ME SNRs are seen within 5$\pm$2 kpc. This coincides with the location of the Molecular Ring seen as a peak in CO emissivity between 3.5 and 7.5 kpc \citep{clemens88}. We note that a correlation with dense molecular material is to be expected as theoretical studies show OH(1720 MHz) masers require densities of order 10$^5$ cm$^{-3}$ \citep{lockett99}.

To test the correlation of ME SNRs with the total distribution of SNRs we apply the Kolmogorov-Smirnov(KS) test. Figure \ref{fig:histograms}b shows the cumulative distribution function of SNRs compared to that of ME SNRs as a function of Galactic latitude. It is immediately clear that ME SNRs are more centrally concentrated; the KS test gives a 99\%\ certainty that ME SNRs do not follow the longitudinal Galactic SNR distribution. Similarly poor correlations are found between ME SNRs and the distribution of both bright SNRs ($\Sigma_{\it 1 GHz} >$5$\times$10$^{-20}$ W m$^{-2}$ Hz sr$^{-1}$) and CO gas. However, when only examine SNRs in the first quadrant, the focus of our new survey, we find that the ME SNRs are not distributed significantly differently from the total SNR population in this region. If current surveys are taken to be complete, then 15\% of SNRs within the Molecular Ring have masers.

Maser-emitting SNRs are among the brightest in the Galaxy \citep{green97}. Figure \ref{fig:histograms}c shows the distribution of surface brightness at 1 GHz for SNRs with masers (filled histogram). The distributions of SNRs which have been searched for masers in previous surveys (Frail et al. 1996; Green et al. 1997) and from this work and given as unfilled and hatched histograms, respectively. All surface brightnesses are taken from Green's catalog (2006). The median surface brightness for ME SNRs is 2.6$\times$10$^{-20}$ W m$^{-2}$ Hz sr$^{-1}$. This is an order of magnitude greater than the median surface brightness for all Galactic SNRs, 3.5$\times$10$^{-21}$ W m$^{-2}$ Hz sr$^{-1}$ \citep[Figure 4]{green04}. 

It is noteworthy that the newly detected ME SNRs are of relatively lower surface brightness with an average of 4.5$\times$10$^{-21}$ W m$^{-2}$ Hz sr$^{-1}$. This work has focused on the region in the inner Galaxy surveyed by \citet{brogan06} where a deep 90cm survey yields a complete census of SNRs in the region down to a surface brightness of $\sim$10$^{-21}$ W m$^{-2}$ Hz sr$^{-1}$. As can be seen in Figure 2c, fewer bright SNRs are included in this survey than in the previous surveys. While ME SNRs are preferentially brighter, there is a large range of surface brightness. As future radio surveys uncover faint remnants in the inner Galaxy it can be expected that a comparable fraction will harbor OH(1720 MHz) masers. 

We also consider whether maser surveys are sufficiently sensitive to detect all ME SNRs. The distribution of peak flux densities for individual masers resolved by interferometric observations is given in Figure \ref{fig:histograms}d. The 5$\sigma$ detection threshold of both single dish surveys(solid line) and interferometer follow-up(dashed line) is indicated. 
\citet{green97} concluded that single dish surveys for OH(1720 MHz) masers were complete as few masers with a peak flux density below 100 mJy had then been detected. However, in the current sample more than a quarter of the detected maser spots have fluxes below 100 mJy, including four of the five masers in this work. Given these updated results, the existing single dish surveys cannot be taken to be complete. 

Interferometric surveys have covered fewer SNRs with higher sensitivity, but it is also unclear whether these are complete. The distribution of maser peak flux densities falls steeply from a peak at the detection threshold of VLA surveys. Many, but not all, deep VLA observations do detect masers down to the sensitivity limits of the observations \citep[for example]{claussen97}. This is suggestive that sensitivity limits the detection of fainter masers. It is not clear how many additional ME SNRs could be detected by deeper surveys.

\section{Conclusion}
We have surveyed 75 known SNRs in the inner Galaxy yielding four new detections of SNRs with OH(1720 MHz) masers: G5.4-1.2, G5.7-0.0, G8.7-0.0 and G9.7-0.0. 
%SNRs G5.7-0.0 and G8.7-0.1 are coincident with sources of TeV $\gamma$-ray emission which may indicate cosmic ray enhancement. 
These four SNRs are clearly interacting with adjacent molecular clouds and are excellent target for further study. 
We present statistical analysis showing ME SNRs are generally of high surface brightness and preferentially distributed in the inner Galaxy where molecular gas is more abundant, as had been previously suggested. We also find that the sensitivity of existing maser surveys may not be sufficient to detect all ME SNRs. Sensitive interferometric observations may determine whether lower flux density masers are common and how profitable a more sensitive survey would be. We note that the four new maser-emitting remnants detected here are all of average surface brightness. As further low surface brightness remnants are detected they are likely to yield further maser detections. 

\acknowledgements
Support for this work was provided by the NSF through award GSSP06-0009 from the NRAO. This Research has made use of NASA's Astrophysics Data System Service and the SIMBAD database, operated at CDS, Strasbourg, France.

\end{document}